\documentclass{article}
\usepackage{url,hyperref,lineno,microtype,bm,graphicx,authblk,amsmath,siunitx,mathtools}
\usepackage[left=2cm, right=2cm, top=2cm, bottom=2.5cm]{geometry}
\usepackage[onehalfspacing]{setspace}

\DeclareSIUnit{\bar}{bar}
\author[a,*]{Olav Galteland}
\author[a]{Michael T. Rauter}
\author[a]{Mina S. Bratvold}
\author[a]{Thuat T. Trinh}
\author[a]{Dick Bedeaux}
\author[a]{Signe Kjelstrup}
\affil[a]{PoreLab, Department of Chemistry, Norwegian University of Science and Technology}
\affil[*]{Corresponding author: olav.galteland@ntnu.no}
\title{Local thermodynamic description of isothermal single-phase flow in porous media}

\begin{document} 
\newcommand{\Def}{\vcentcolon =}
\maketitle
\begin{abstract}
Darcy's law for porous media transport is given a new local thermodynamic basis in terms of the grand potential of confined fluids. The local effective pressure gradient is determined using non-equilibrium molecular dynamics, and the hydraulic conductivity and permeability are investigated. The transport coefficients are determined for single-phase flow in face-centered cubic lattices of solid spheres. The porosity changed from that in the closest packing of spheres to near unity in a pure fluid, while the fluid mass density varied from that of a dilute gas to a dense liquid. The permeability varied between \SI{5.7e-20}{\meter^2} and \SI{5.5e-17}{\meter^2}, showing a porosity-dependent Klinkenberg effect. Both transport coefficients depended on the average fluid mass density and porosity but in different ways. These results set the stage for a non-equilibrium thermodynamic investigation of coupled transport of multi-phase fluids in complex media.
\end{abstract}

\maketitle

\section{Introduction}
Porous media are everywhere in nature and technology, and transport through them is important. To take some widely different examples; we need to describe the transport of nanoparticles across cell layers with medicine to cancerous tissue \cite{Stylianopoulos2018}. We also need to describe the selective transport across the porous separators in batteries and fuel cells \cite{Zlotorowicz2015}. We are aiming for a description that reflects the underlying properties of the single pores. 

Much of the fundamental work on flow in porous media has been done for the pore-scale, see Helmig for a pedagogical presentation \cite{Helmig1997}. Recent developments in imaging techniques and computer capabilities have improved our understanding of the physics at pore-scale considerably \cite{blunt2017multiphase}. But there is no consensus on how to upscale from the pore-scale to the Darcy scale. On the Darcy scale, transport is described as taking place between representative elementary volumes (REVs).

\begin{figure}
    \centering
    \includegraphics[width=0.5\linewidth]{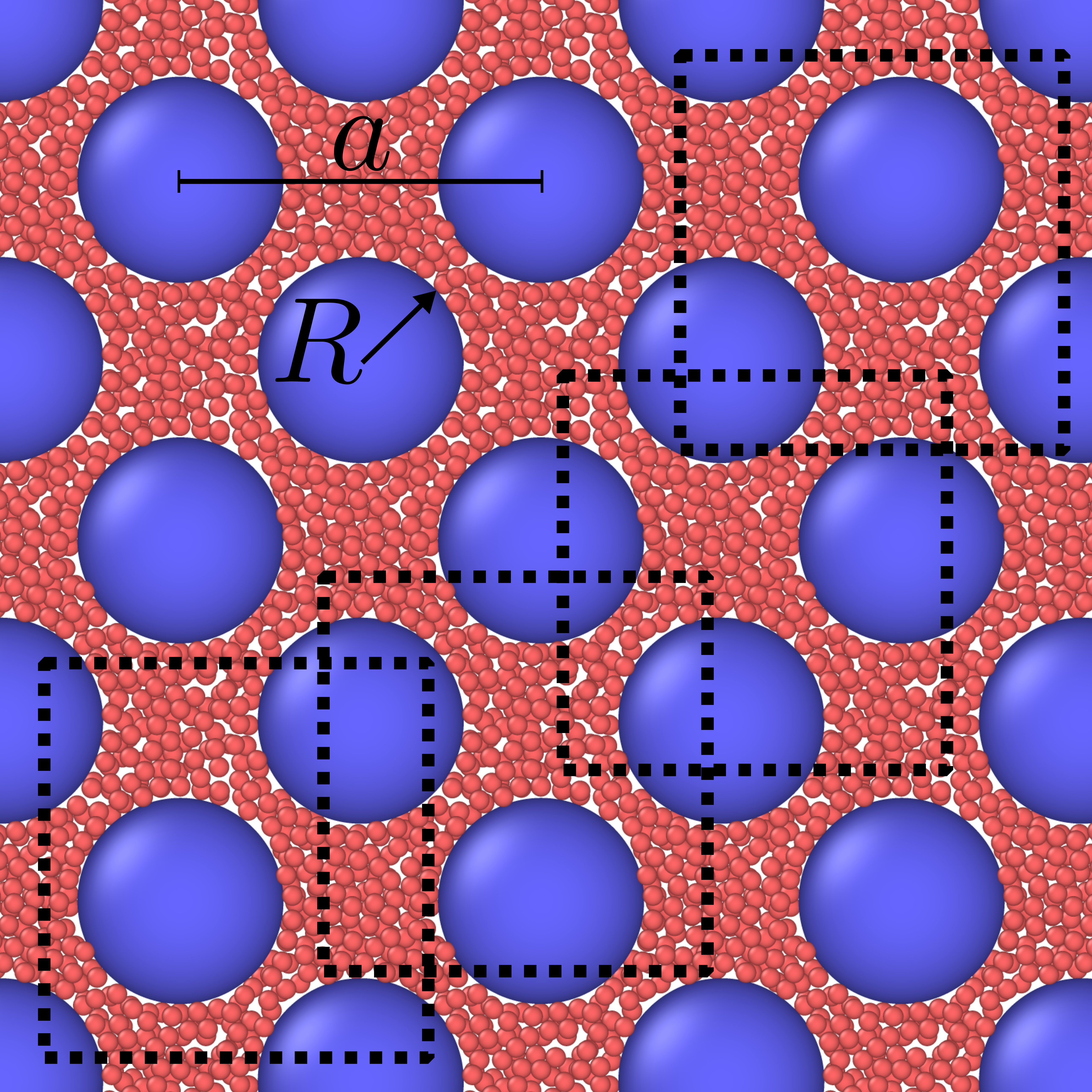}
    \caption{The representative elementary volume (REV) for a single-phase fluid in a face-centered cubic lattice of solid spheres is the same size as the unit cell of the lattice. The radius of the solid particles is $R$ and the lattice constant is $a$. The squares with dashed lines mark the magnitude of the REV. All regions have the same thermodynamic properties.}
    \label{fig:REV}
\end{figure}

The REV in this work is defined to be large enough to be statistically representative of the system. From a statistical mechanics point of view, the REV includes all available microstates of the system. In this work, we are investigating a compressible single-phase fluid in a porous structure made up of solid particles in a face-centered cubic (fcc) lattice. The system is illustrated in Fig. \ref{fig:REV}. The blue large spheres represent solid, while the red particles represent the fluid. We shall use additive variables to define the REV, similar to Whitaker \cite{whitaker1986flow}. This definition of the REV differs from the one suggested by Nordahl and Ringrose \cite{nordahl2008identifying} who used a constant permeability as criterion. Because of the fcc symmetry, a unit cell is here a proper choice of REV. Around any point in the porous medium, we can choose a unit cell with this point at its center. In this way, we can obtain a continuous path between unit cells on the REV scale.
We first determine the pressures of the REV at equilibrium, as a function of the temperature, fluid mass density, and porosity, $\hat{p}(T,\rho_f,\phi)$. This can be regarded as finding the equation of state of the REV. Up to this point, we have used REV densities much like Whitaker \cite{whitaker1986flow}.  

Thermodynamic theories for transport in porous media are little developed. In confinement, fluid thermodynamic properties will deviate from their corresponding bulk phase values. In an attempt to find a continuous description on the Darcy scale, we need to reflect on properties of the pore scale or sub-pore scale, including the nanometer scale. Hence, we need descriptions of fluxes and forces for the REV of the porous medium. Is it possible to find a description that does not explode in complexity, but still brings forward the characteristic properties of the smaller scale, such as wetting and adsorption? Pore-scale descriptions of REVs have contained up to 27 variables \cite{Gray1998}. We have claimed that we can reduce the number of variables to a more practical number, and describe here the first steps in the direction to apply this relatively new thermodynamic procedure for coarse-graining \cite{Galteland2019, Rauter2020, bedeaux2021fluctuation}. 

In our search for a thermodynamic coarse-graining procedure to be applied to transport in porous media, we have chosen to define effective thermodynamic variables by sets that combine in an additive manner \cite{Kjelstrup2018, Kjelstrup2019, bedeaux2021fluctuation}. We have derived the Gibbs equation for the REV, and in the analysis of transport, we have assumed validity of Gibbs equation. This gives the entropy production for the REV and a thermodynamic basis for Darcy's law in isothermal system. Our proposal to find the equation of state and the Gibbs equation is new. Here, we apply the method to isothermal, compressible, single-phase flow. This entails obtaining the REV grand potential as the sum of contributions from all phases and surfaces. A theory of transport on the Darcy scale follows in a way which is standard in non-equilibrium thermodynamics. 

\begin{figure}
    \centering
    \includegraphics[width=\linewidth]{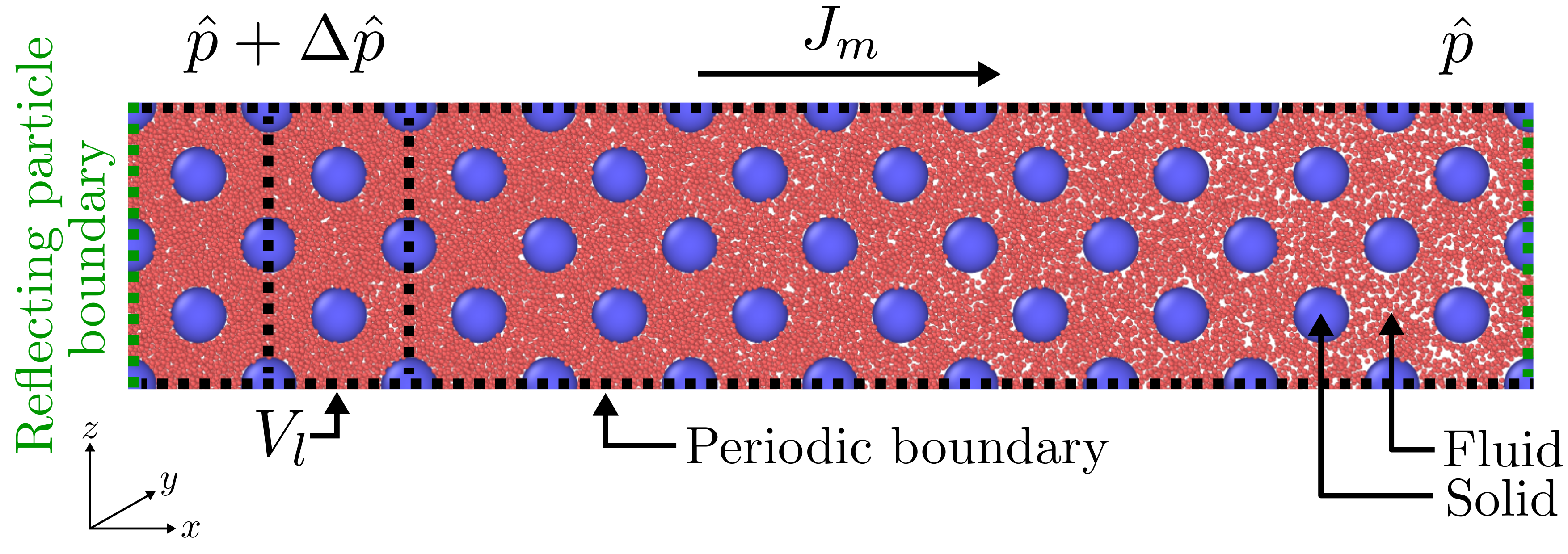}
    \caption{A visualization of a simulation, with porosity $\phi\approx 0.87$ and average bulk fluid mass density $\rho_b=\SI{237(1)}{\kilo\gram\per\meter^3}$. The blue particles represent the solid particles of a fcc lattice, while the red particles represent the fluid particles. There is an integral pressure difference $\Delta\hat{p}$, which drives a mass flux $J_m$ from left to right. The slab of volume $V_l$ was used as REV.  The simulation box had periodic boundaries in all directions (black dashed line), apart from particles crossing the $x$-boundary from left to right. Here the reflecting particle method was applied, see section \ref{sec:sim} for details of this boundary condition \cite{li1998coupling}. The simulated system was visualized with OVITO \cite{Stukowski2009}.}
    \label{fig:system}
\end{figure}

An excellent tool to analyze transport in porous media is non-equilibrium molecular dynamics (NEMD) \cite{todd2017nonequilibrium}. This tool allows us to simulate molecular properties (like velocities and forces), yet upscale to fluid properties (like pressure) on the Darcy scale. In NEMD, we solve Newton equations for particles, so the outcome can also be used to assess assumptions made in the thermodynamic theory. In this sense the tool supplements lattice Boltzmann simulations and numerical solutions to the Navier-Stokes equation. A downside is that the length and time scale becomes limited as NEMD is a computationally expensive technique. Here, we will use NEMD to simulate the flow of methane-like molecules in a face-centered cubic (fcc) lattice made up of spherical solid particles, see Fig. \ref{fig:system}. We will investigate a vast range of fluid densities, which varies from highly compressible to nearly incompressible. NEMD has been used to simulate many transport processes in heterogeneous media, as documented by, for example, Ikeshoji and Hafskjold \cite{hafskjold1993molecular, ikeshoji1994non, hafskjold1996non}.

We use our coarse-graining procedure to obtain the effective pressure of the REV. Its negative gradient is the driving force for fluid flow. The effective pressure of a REV, $\hat{p}$, is called the integral pressure. It is in general a combination of pressures, surface tensions, and line tensions. This approach originates from nanothermodynamics, as described by Hill \cite{Hill1994}. We believe, however, that this procedure for other porous media and other fluids. We have shown that the method can replace the use of Young or Young-Laplace's law \cite{Rauter2020}. For a recent formulation of nanothermodynamics, see Bedeaux, Kjelstrup, and Schnell \cite{Bedeaux2020}.

It is our long-range aim to obtain a procedure that provides equations of transport in porous media in general. Not only for isothermal transport of one fluid but also coupled transport due to other driving forces, for example, thermal driving forces \cite{rauter2021thermo,rauter2021cassie}. We start with a single-phase flow, to document the use of new concepts on porous media pressure in a simple way. We have reported equilibrium studies with this coarse-graining procedure previously \cite{Galteland2019,Erdos2020, Rauter2020, Galteland2021, galteland2021legendre, galteland2022defining}. We expand these results to non-equilibrium conditions in this work. 

This work aims to analyze this simple transport problem and compare its results to expressions that are common in the literature, most importantly Darcy's law in the presence of the Klinkenberg effect \cite{Helmig1997}, but also the Kozeny-Carman equation \cite{kozeny1927uber, carman1937fluid, carman1956flow, berg2014permeability}.

Fig. \ref{fig:system} illustrates the chosen system, a fcc lattice of solid particles (blue), all with radius $R$, and a lattice constant $a$. The porosity $\phi$ is varied by varying $a$. The lattice is periodic in all directions. The REV is the size of the lattice unit cell \cite{Galteland2019}, but here we chose to integrate to a slab of volume $V_l$. The volume $V_l$ contains four unit cells and is practical for the purpose. A pressure difference arises using a reflecting particle boundary, which forces a mass flux $J_m$ through the cross-sectional area. This boundary-driven method gives minimal disturbance of fluid particles away from the boundary.

In Section \ref{sec:theory}, we recapitulate the necessary thermodynamic equilibrium relations of the pertinent REV, followed by a description of transport, and a procedure to find pressure profiles away from equilibrium. The equilibrium and non-equilibrium simulation procedures that are applied, are described in Section \ref{sec:sim}. The results are discussed in Section \ref{sec:results}. We give the results needed to determine the local driving force inside the porous medium. The equation of state of the porous medium is central here. The hydraulic conductivity and the permeability are however strong functions of porosity and average fluid mass density. We shall also see that the Klinkenberg correction, as described by Helmig \cite{Helmig1997}, applies to the fluid permeability. 

\section{The thermodynamic variables of the REV}
\label{sec:theory}
\subsection{The grand potential and the pressure}

We have chosen to describe the state of a porous medium using a basis set of thermodynamic variables that are additive in the sense that each coarse-grained variable in the REV is a sum of contributions from all bulk phases, surfaces, and possible contact lines \cite{Kjelstrup2018, Kjelstrup2019, bedeaux2021fluctuation}. The REV variables are controlled by the environment \cite{Bedeaux2020}. In the present case, the control variables are temperature, volume, and fluid chemical potential. The REV of interest belongs therefore to a grand canonical ensemble. We recapitulate the thermodynamic description of a REV of this kind at equilibrium before we define the situation in the presence of flow. 

In nanothermodynamic theory, the grand potential $\Upsilon$ of the REV is given by minus the so-called \textit{integral} pressure of the REV times the volume. The adjective ``integral" was coined by Hill \cite{Hill1994} to reflect the fact that it is an integrated property. Conversely, the normal pressure was called the differential pressure. We can regard the integral pressure as the \textit{effective} pressure of the REV. The grand potential is,
\begin{equation}
    \Upsilon =-\hat{p}V.
    \label{eq:grand_potential}
\end{equation}
In this expression, $\hat{p}$ and $V$ are the integral pressure and the volume of the REV, respectively. It can be understood as the defining equation for $\hat{p}$. The grand potential has its basis in statistical mechanics
\begin{equation}
    \Upsilon =-k_{\text{B}}T\ln \Xi,
    \label{eq:ln}
\end{equation}
where $\Xi$ is the grand canonical partition function, $k_\text{B}$ is Boltzmann's constant, and $T$ is the temperature. This basis explains why $\Upsilon$ has additive contributions. Weakly coupled sub-systems will add to $\Upsilon$, and result in a product of partition functions. This expression explains also why the REV needs to include all possible micro-states of the system.

A more colloquial name of the grand potential is the compressional energy, since it is a product of pressure and volume, which is related to work. In the present case, the grand potential has several additive contributions. We obtain
\begin{equation}
    \hat{p}V = \hat{p}_fV_f + \hat{p}_sV_s - \hat{\gamma} A
    \label{eq:grand}
\end{equation}
The contributions are from the fluid and solid phases, and the fluid-solid surface. The symbols $\hat{p}_f$, $\hat{p}_s$ are the integral pressure of the fluid and solid, respectively, and $\hat{\gamma}$ is the integral fluid-solid surface tension. The surface energies are more significant because of the fluid confinement. For large fluid volumes, the surface energy may be neglected. However, when Young-Laplace's and Young's equations are significant for multi-phase flow the surface energies in this equation are too. The volume of the fluid and solid are $V_f$ and $V_s$, respectively, and $A$ is the surface area between the solid and the fluid. The sum of the fluid and solid volumes is equal to the REV volume $V=V_f+V_s$, and the porosity is $\phi=V_f/V$. It follows that the fluid mass density of the REV is
\begin{equation}
    \rho_f = \frac{M_f}{V_f+V_s}
    \label{eq:dens}
\end{equation}
where $M_f$ is the mass of the fluid in the REV. This fluid mass density follows the coarse-graining procedure. This density is not the internal density given by $M_f/V_f$. The individual contributions to the integral pressure are not needed, we only need the total sum. The integral pressure is obtained by calculating an equation of state in equilibrium conditions, which is applied to non-equilibrium conditions. The equation of state for the porous medium must as all equations of state be found from experiments or simulations. In the present work, the last method gives the equation of state for the present medium
\begin{equation}
    \hat{p} = \hat{p}(T, \rho_f,\phi).
    \label{eq:EoS}
\end{equation}
The functional dependence of $\hat{p}$ on the variables ($\rho_f,\phi$) will here be investigated by simulations at isothermal conditions. The value of $\hat{p}$ shall also be found by dividing both sides of Eq. \ref{eq:grand} by the REV volume. This gives
\begin{equation}
    \hat{p} = \hat{p}_f\phi + \hat{p}_s(1-\phi) - \hat{\gamma} A/V.
    \label{eq:equi}
\end{equation}
In earlier work we determined $\hat{p}_s$ from known values of $A, V$ and $\phi$ \cite{Galteland2019}. Here we shall use Eq. \ref{eq:equi} to compute $\hat{p}$ and compare the result to the result from Eq. \ref{eq:EoS}. Both procedures are completely general. In Eq. \ref{eq:equi} we apply an independent computation of the single parameters. The two expressions will be shown here to give the same result. 

The general relation between the differential and integral pressures is given by 
\begin{equation}
    p = \left[\frac{\partial (\hat{p}V)}{\partial V} \right ]_{T, \mu} = \hat{p} + V \left(\frac{\partial \hat{p}}{\partial V}\right)_{T,\mu}
    \label{eq:diff_int}
\end{equation}
This shows that $p$ or $\hat{p}$ can enter the equation of state for the porous medium, see Eq. \ref{eq:EoS}. The fcc lattice used in the present model poses a special condition on the two pressures. Because of the lattice symmetry, the integral pressure will not depend on the REV volume, giving the special condition $p=\hat{p}$ for the REV. Any block of adjacent unit cells will therefore also give the same value of $\hat{p}$, see Fig. \ref{fig:REV}. It follows that the integral pressure in this lattice is independent of the size of the REV. As a consequence, the differential pressure of the REV is identical to the integral pressure, $p = \hat{p}$ for the REV. This is only the case for this special system, and not in general. To be general, we will keep the integral pressure $\hat{p}$.

Such an equality does \textit{not} apply for $\hat{p}_f$, $ \hat{p}_s$ and $\hat{\gamma}$, which in general differ from $p_f, p_s$ and $\gamma$. Only if the spheres are far apart, $\hat{p}_f = p_f$, and if the spheres are large enough, also $\hat{\gamma}= \gamma$. This was the special case considered before \cite{Galteland2019}. In that work, we required that the fluid volume was so large that the fluid pressure was equal to the bulk pressure and that the fluid-solid curvature was so small that the surface tension did not depend on it. By using such requirements, we cannot take into account any disjoining pressure \cite{israelachvili2015intermolecular} or Tolman length \cite{tolman1949effect}. The procedure in this work is more general and can describe such effects and other capillary effects. 

The presence of additional fluid phases would contribute by similar terms to Eq. \ref{eq:equi}. The saturation will appear as a variable, and there are contributions from three-phase contact lines. Eq. \ref{eq:equi} is an alternative definition of the integral pressure of the REV. When there is equilibrium at the boundary between the porous medium and the bulk phase surrounding the porous medium, we have the condition 
\begin{equation}
    \hat{p} = p_b.
   \label{eq:bulk_pressure}
\end{equation}
This condition was recently used to determine the solid integral pressure $\hat{p}_s$ \cite{Galteland2021}.  We shall here use the relation \ref{eq:bulk_pressure} to determine the REV integral pressure in the equation of state (Eq. \ref{eq:EoS}). The equation of state is next applied to non-equilibrium conditions assuming local equilibrium. This procedure is possible for any geometry. 

\subsection{The effective pressure gradient}

The Gibbs equation can be written in terms of coarse-grained variables of the REV, as described in the previous section. By introducing the entropy, mass, and energy balance equations, we can then obtain the entropy production of the REV. In the present isothermal single-phase case, the entropy production has only one flux-force product. The flux conjugate to the negative pressure gradient is the volume flux, $J_V$
\begin{equation}
    J_V = - l \frac{\partial \hat{p}}{\partial x}.
    \label{eq:NET}
\end{equation}
This expression gives a \textit{thermodynamic} basis for Darcy's law; which is a locally linear relationship between the volume flux and the driving force valid for the porous medium. According to non-equilibrium thermodynamics, this equation does not assume laminar flow conditions. The only assumption is that there is local equilibrium and that the fluxes are linear combinations of the forces. These assumptions have been shown to hold for coupled heat and mass transport through liquid-vapor interfaces \cite{rosjorde2000nonequilibrium} and membranes \cite{inzoli2009transport}. The conductivity coefficient (the hydraulic conductivity), $l$, is a function of state variables. The driving force is not necessarily constant along the medium. We need the local value of $\partial \hat{p}/\partial x$ and $J_V$ to find $l$.  

When the porous medium is in contact with bulk fluids at both ends, the boundary condition will always give $\Delta \hat{p} = \Delta p_b$. In the experimental situation, the gradient is the pressure difference divided by the length $L$ of the medium and the conductivity coefficient is the coefficient $l$ obtained from
\begin{equation}
    J_V = - l \frac{\Delta \hat{p}}{L}  = - l \frac{\Delta p_b}{L}.
\end{equation}
The permeability $k$ enters Darcy's law on the form
\begin{equation}
    J_V = - \frac{k}{\eta_b} \frac{\Delta \hat{p}}{L} =  -  \frac{k}{\eta_b} \frac{\Delta {p}_b}{L},
    \label{eq:Darcy}
\end{equation}
where $\eta_b$ is the shear viscosity of the bulk fluid, and 
\begin{equation}
    l  = \frac{k}{\eta_b}
\end{equation}

The volume flux is related to the mass flux by the fluid mass density, $J_V = J_m/\rho_f$. While $J_V$ varies, $J_m$ is constant in the steady state, 
\begin{equation}
    J_m = {J_V}{\rho_f} = - \rho_f \frac{k}{\eta_b} \frac{\partial \hat{p}}{\partial x}.
    \label{eq:mass_flux}
\end{equation}
We shall determine the coefficients $l$ and $k$ from the last two equations, with information of the mass flux $J_m$, fluid mass density of the REV, $\rho_f$, and integral pressure gradient $\partial \hat{p}/\partial x$. Their dependence on the fluid mass density $\rho_f$ and the porosity $\phi$ is of interest.

The original form of Darcy's law was obtained experimentally for the laminar flow of a single-phase fluid through a porous medium. It has also been given a mechanical basis \cite{Helmig1997}. The Reynolds number, used to characterize flow patterns, is here calculated from $\text{Re} = 2RJ_m/\eta_b$, where $J_m$ is the mass flux, and $2R$ the diameter of the solid particles. For Reynolds numbers considered in this work, the flow is laminar \cite{Helmig1997}.

A question of general interest is how the permeability of a porous medium, as used in Darcy's law, can be related to the properties of the porous medium. A well-known approach for a bed of granular solids is given by the Kozeny-Carman equation. As the name suggests, it was first derived by Kozeny \cite{kozeny1927uber}, assuming that the fluid in the porous medium could be described as contained in a set of non-interfering parallel channels with the same internal surface and pore volume, as the medium itself. The solid phase was a packed bed. The general equation was written as \cite{kozeny1927uber,carman1937fluid, carman1956flow, berg2014permeability}
\begin{equation}
    J_V = - \frac{\tau^2V^2\phi^3}{c_0 A^2} \frac{1}{\eta_b}\frac{\Delta p}{L}
    \label{eq:Kozeny}
\end{equation}
where $\phi$ is the porosity, $c_0$ denotes the Kozeny constant, $\tau$ is the tortuosity, and $A/V$ is the surface-to-volume ratio. The tortuosity is the average length a particle travels across the medium divided by the medium length $L$. For monodisperse spheres of diameter $d$ the permeability simplifies to
\begin{equation}
    k = \frac{\tau^2d^2}{c_0} \frac{\phi^3}{(1-\phi)^2}.
\end{equation}
Klinkenberg gave a correction for low-density compressible fluids with basis in fluid slippage at the wall \cite{Klinkenberg1941}. When this correction is included, the effective permeability, $k$, is written as
\begin{equation}
   k = k_0 \left(1 + \frac{b}{p}\right)
\end{equation}
where $k_0$ is the absolute permeability and $b$ is the Klinkenberg constant \cite{Helmig1997}.  The correction was found to be relevant for permeabilities $ k $< 10 $^{-13}$ m$^2$ \cite{Baehr1991}.

\section{Molecular simulations}
\label{sec:sim}
\subsection{The porous medium}
Systems of a single-phase and single-component fluid were simulated in a porous structure using molecular dynamics simulations with LAMMPS \cite{thompson2021lammps}. The porous solid structure was composed of solid spherical particles in a face-centered cubic (fcc) lattice. The fluid particles were free to move, while the solid particles were immovable. As a consequence, the porous medium was non-deformable. The mass of the fluid particles was $m$, while the mass of the solid particle was not defined since they were immovable.  The radius of the solid particles was $R$ and the lattice constant was $a$. 

The particles interacted with the Lennard-Jones/spline potential, which is a pair-wise interaction potential that models non-bonded neutral atom interactions. It is for example an accurate model for noble gasses and methane \cite{steele1974interaction, miyahara1997freezing, mcgaughey2004thermal, rutkai2017well}. See Hafskjold \textit{et al.} \cite{Hafskjold2019} and Kristiansen \cite{kristiansen2020transport} for details on the thermodynamic and transport properties for this potential. The potential is
\begin{equation}
    u(r) = 
    \begin{cases}
    \infty & \text{if } r \leq d,\\
    4\epsilon\left[\left(\frac{\sigma}{r-d}\right)^{12}-\left(\frac{\sigma}{r-d}\right)^{6}\right] &\text{if } d < r \leq r_s,\\
    a(r-r_c)^{2}+b(r-r_c)^{3} &\text{if } r_s<r\leq r_c,\\
    0 &\text{if } r>r_c.
    \end{cases}
\end{equation}
Where $r= \|\bm{r}_j-\bm{r}_i\|$ was the distance between particle $i$ and $j$. The interaction strength was characterized by the depth of the interaction potential $\epsilon$, and the particle diameter was characterized by the distance $\sigma$. The hard-core diameter of the particles was $d$. The parameters $a, b, r_s$ and $r_c$ were set such that the potential and its derivative were continuous at $r_s$ and $r_c$.

The simulation data were related to SI units by using parameters $\epsilon$, $\sigma$, and the fluid particle mass $m$ for methane \cite{steele1974interaction, miyahara1997freezing}. The aim of this is not to give accurate data for methane transport, but rather to present more identifiable values. The parameters were in terms of SI units
\begin{equation}
    \epsilon/k_\text{B} =  \SI{148.1}{\kelvin}, \quad \sigma= \SI{0.381}{\nano\meter}, \quad\text{and}\quad m =  \SI{2.661e-26}{\kilo\gram}.
\end{equation}
The parameters $\epsilon_{\text{ff}}=\epsilon_{\text{fs}}=\epsilon$ and $\sigma_{\text{ff}}=\sigma_{\text{fs}}=\sigma$ were equal for the fluid-fluid and solid-fluid interactions, while the hard-core diameter was zero for the fluid-fluid interactions $d_{\text{ff}}=0$ and $d_{\text{fs}}=4.5\sigma\approx\SI{1.7}{\nano\meter}$ for the fluid-solid interactions. The solid did not interact with other solid particles. The radius of the solid was constant and defined to be $R \Def d_{\text{fs}}+\sigma_{\text{fs}}/2=5\sigma\approx\SI{1.9}{\nano\meter}$, at which point the potential energy of the fluid-solid interaction is zero. The fluid-solid slip conditions were not precisely defined, but depended on the shape of the fluid-solid surface and fluid-solid interactions. The fluid-solid surface was completely smooth.

The lattice constant was varied from $a=2\sqrt{2}R\approx \SI{5.4}{\nano\meter}$ to $a=40\sigma\approx\SI{15.2}{\nano\meter}$, where the lower limit is where the surface of the solid particles are in contact. The porosity and surface-to-volume ratio of the porous structure were
\begin{equation}
   \phi = 1 - \frac{16\pi R^3}{3a^3}\quad\text{and}\quad A/V=\frac{16\pi R^2}{a^3}.
\end{equation}
Consequently, the porosity varied from $0.26$ to $0.97$ and the surface-to-volume ratio varied between approximately \SI{0.05}{\per\nano\meter} and \SI{1.17}{\per\nano\meter}. The fcc lattice has four octahedral and eight tetrahedral voids. The octahedral voids are the largest and can accommodate a sphere of radius $a/2-R$, which varies from \SI{0.8}{\nano\meter} to \SI{5.7}{\nano\meter} in the cases studied here.

\subsection{Shear viscosity}
\begin{figure}
    \centering
    \includegraphics[width=0.8\linewidth]{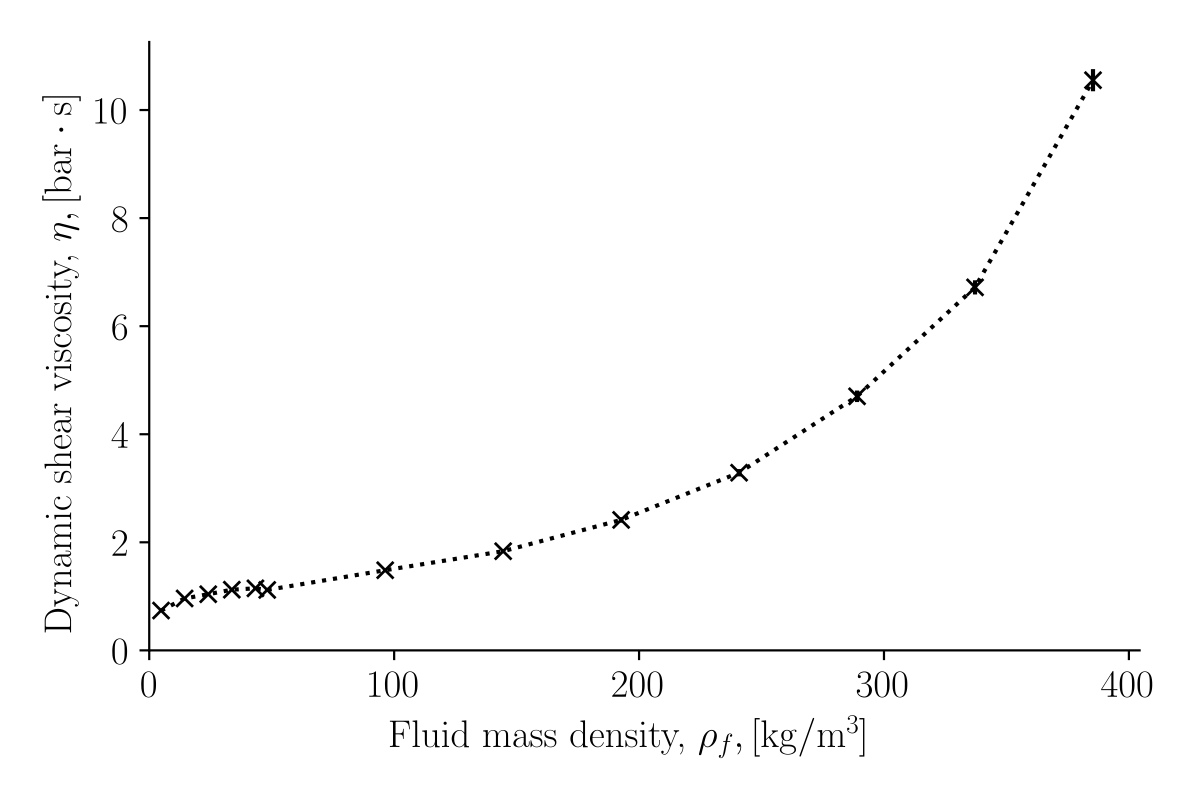}
    \caption{Shear viscosity as a function of fluid mass density.}
    \label{fig:viscosity}
\end{figure}

The shear viscosity of the bulk fluid was calculated with the OCTP plugin \cite{Jamali2019} from the time-integral of the auto-correlation function of the off-diagonal components of the Cartesian mechanical pressure tensor,
\begin{equation}
    \eta_{\alpha \beta}=\lim _{t \rightarrow \infty} \frac{1}{2 t} \frac{V}{k_{B} T}\left\langle\left(\int_{0}^{t} P_{\alpha \beta}\left(t^{\prime}\right) \mathrm{d} t^{\prime}\right)^{2}\right\rangle.
\end{equation}
The system consisted of 32000 fluid particles and was initialized at temperature $T = \SI{296.2}{\kelvin}$ and bulk mass density in the range $4.8$ to \SI{385}{\kilo\gram\per\meter^3}. The system was equilibrated with an $NVT$-ensemble for $2\times10^6$ steps. After that, an additional $2\times10^6$ steps were run in the $NVE$ ensemble to collect data and compute the shear viscosity. The time step was $\delta t = \SI{1.375}{\femto\second}$. The statistics were improved by running 30 independent simulations. The shear viscosity is shown as a function of the fluid mass density in Fig. \ref{fig:viscosity}.

\subsection{Simulation procedure}
\label{sec:simproc}
\subsubsection{Equilibrium conditions}
The systems were initialized by creating a fcc unit cell of solid particles with varying lattice constant $a$ in a cubic simulation box of side lengths $a$. In addition, a bulk fluid without solid particles was simulated. The boundaries of the simulation box were periodic. The temperature of the fluid was controlled with a Nosé-Hoover type thermostat to be constant and equal to $T=2\epsilon/k_\text{B}=\SI{296.2}{\kelvin}$ \cite{Shinoda2004}, which is in the supercritical region ($T_c=0.885\epsilon/k_\text{B}\approx \SI{131}{\kelvin}$  \cite{Hafskjold2019}). A supercritical fluid was investigated to avoid phase separation.

Fluid particles were inserted and removed from the porous structure using the grand canonical Monte Carlo \cite{frenkel2002computational}. This was done to generate initial configurations with varying lattice constants that were in equilibrium with each other. Fluid particles were inserted into the pores with an acceptance probability
\begin{equation}
    \text{acc}(N_f\rightarrow N_f+1) = \min\left\{1,\quad \frac{V}{\Lambda^3(N_f+1)}\exp[\beta(\mu_f-\Delta E_p] \right\}
\end{equation}
and a random fluid particle was removed from the simulation box with an acceptance probability
\begin{equation}
    \text{acc}(N_f\rightarrow N_f-1) = \min\left\{1,\quad \frac{\Lambda^3(N_f+1)}{V}\exp[-\beta(\mu_f+\Delta E_p] \right\}.
\end{equation}
Where $\beta=1/k_\text{B}T$ is the thermodynamic beta, $\Lambda=\sqrt{h^2/(2\pi m k_\text{B}T)}$ is the de Broglie thermal wavelength, $h$ is the Planck constant, $N_f$ the number of fluid particles, and $\Delta E_p$ is the potential energy difference of the system, when inserting or removing a fluid particle. The fluid chemical potential varied in the interval $\mu_f\in[-10, 10]\epsilon$, which resulted in a bulk fluid mass density variation from $\rho_b=\SI{3.3(2)}{\kilo\gram\per\meter^3}$ to $\SI{409.9(5)}{\kilo\gram\per\meter^3}$. The simulations were run until the average number of fluid particles reached a constant value. For the largest porosities and chemical potentials, this took up to $9\times10^6$ steps. The time step was $\delta t = \SI{2.75}{\femto\second}$.

\begin{figure}
    \centering
    \includegraphics[width=0.6\linewidth]{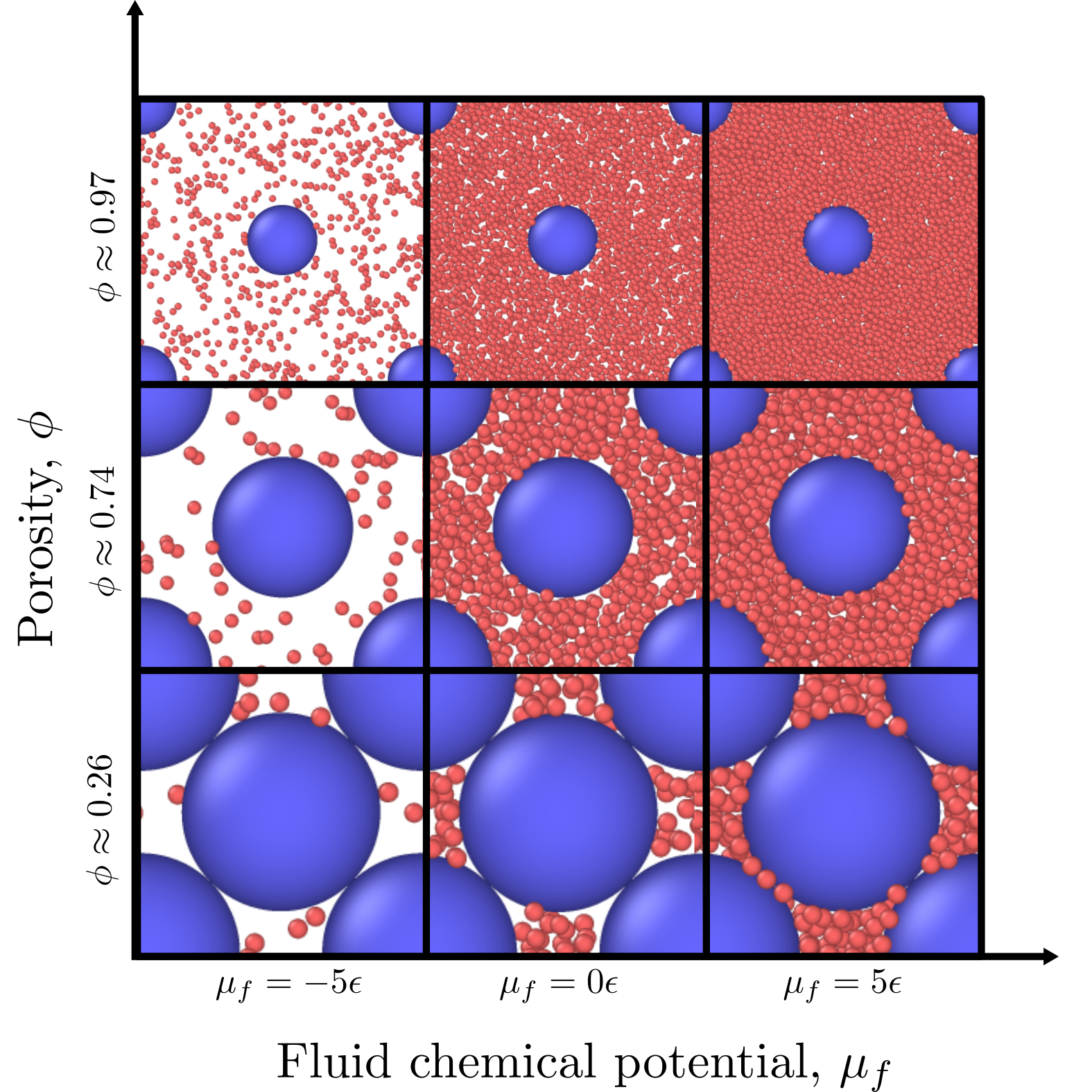}
    \caption{Nine initialized simulations in equilibrium conditions of varying fluid chemical potential and porosity. The three chemical potentials used in the snapshots correspond to bulk mass densities $\rho_b=[(39.9\pm0.8), (236.0\pm1.1), (350.9\pm0.8)] \SI{}{\kilo\gram\per\meter^3}$ from left to right. Each illustration is scaled such that the figure sizes become the same. In the simulations, the radius of the blue solid particles was constant in all simulations. The simulated system was visualized with OVITO \cite{Stukowski2009}.}
    \label{fig:porosity_chemical_potential}
\end{figure}

Fig. \ref{fig:porosity_chemical_potential} illustrates nine visualizations of the equilibrium simulations for three porosities and three chemical potentials. The figure provides a visual impression of the range of conditions studied, from closest packing of spheres and porosity $\phi\approx 0.26$ (bottom row) to an open fcc lattice with porosity $\phi\approx 0.97$ (top row). The three chemical potentials used in the snapshots correspond to bulk mass densities $\rho_b=[(39.9\pm0.8), (236\pm1), (350.9\pm0.8)] \SI{}{\kilo\gram\per\meter^3}$ from left to right. The figure parts are scaled such that the figure sizes become the same. In the simulations, the radius of all blue solid particles was the same. 

\subsubsection{Non-equilibrium conditions}
The unit cells were replicated twice in the $y$- and $z$-directions and ten times in the $x$-direction in these studies. The final side lengths of the simulation box in non-equilibrium conditions were $(10, 2, 2)a$, where $a$ is the lattice constant. The grand canonical Monte Carlo fluid particle insertions and removals from the porous medium were stopped, while the temperature was controlled with a Nosé-Hoover type thermostat \cite{Shinoda2004}. However, the temperature was also controlled separately in each REV (volume $V_l$) to ensure that there was no temperature gradient.

The reflecting particle method (RPM) \cite{li1998coupling} was applied to the $x$-boundary of the simulation box to induce an integral pressure gradient. The RPM allowed particles to cross the periodic $x$-boundary from right to left with a probability $1-q$ and be reflected with a probability $q\in[0,1]$. A value $q=0$ entailed that no fluid particles were reflected, while a value $q=1$ entailed that all fluid particles were reflected when attempting to cross the $x$-boundary from right to left. The fluid particles were never reflected when crossing the $x$-boundary from left to right. The pressure gradient was controlled by varying the probability $q$. This is a boundary-driven non-equilibrium molecular dynamics method to induce a pressure gradient that has minimal disturbance on the fluid particles away from the boundary. The simulations were run until they had reached steady-state, where the mass flux was constant along the $x$-axis. For the longest simulations, this took up to $8\times10^6$ steps. The time step was $\delta t =\SI{ 2.75}{\femto\second}$.

The mass fluxes and integral pressures were calculated in layers $l$ of volume $V_l$ along the $x$-axis. The volume of each layer was equal to the lattice constant $a$ and spanned the simulation box in the $y$- and $z$-directions. The side lengths of the layers were $(a, 2a, 2a)$ in the $x$-, $y$- and $z$-directions, respectively. The volume of each layer (the REV) was consequently $V_l=4a^3$. Each layer contains four unit cells. The mass flux through layer $l$ was calculated as
\begin{equation}
    J_{l,m} = \frac{1}{V_l}\sum_{i\in V_l} m_iv_{i,x} = \rho_{f,l}\langle v_{i,x}\rangle
\end{equation}
where $m_i$ is the mass of fluid particle $i$, $v_{i,x}$ is the velocity of fluid particle $i$ in the $x$-direction, and $\langle v_{i,x}\rangle$ is the average particle velocity in the $x$-direction. The sum is over all particles in the layer $V_l$.  The integral pressure of each layer was calculated by relating the fluid mass density $\rho_{f,l}$ to the integral pressure calculated in equilibrium.

With this procedure, we were able to determine the transport coefficients $l$ and $k$. The procedure has been used earlier by us to establish the same gradient in integral pressure \cite{galteland2022defining}, however, the permeability calculations are done for the first time here.

\section{Results and discussion}
\label{sec:results}
The results of this work are shown in Figs. \ref{fig:fluid_density} - \ref{fig:klinkenberg}. The results are presented and discussed with reference to the simulation procedure (Section \ref{sec:simproc}) and to the Theory (Section \ref{sec:theory}).

\subsection{Equilibrium conditions. The equation of state}

\begin{figure}
    \centering
    \includegraphics[width=0.8\linewidth]{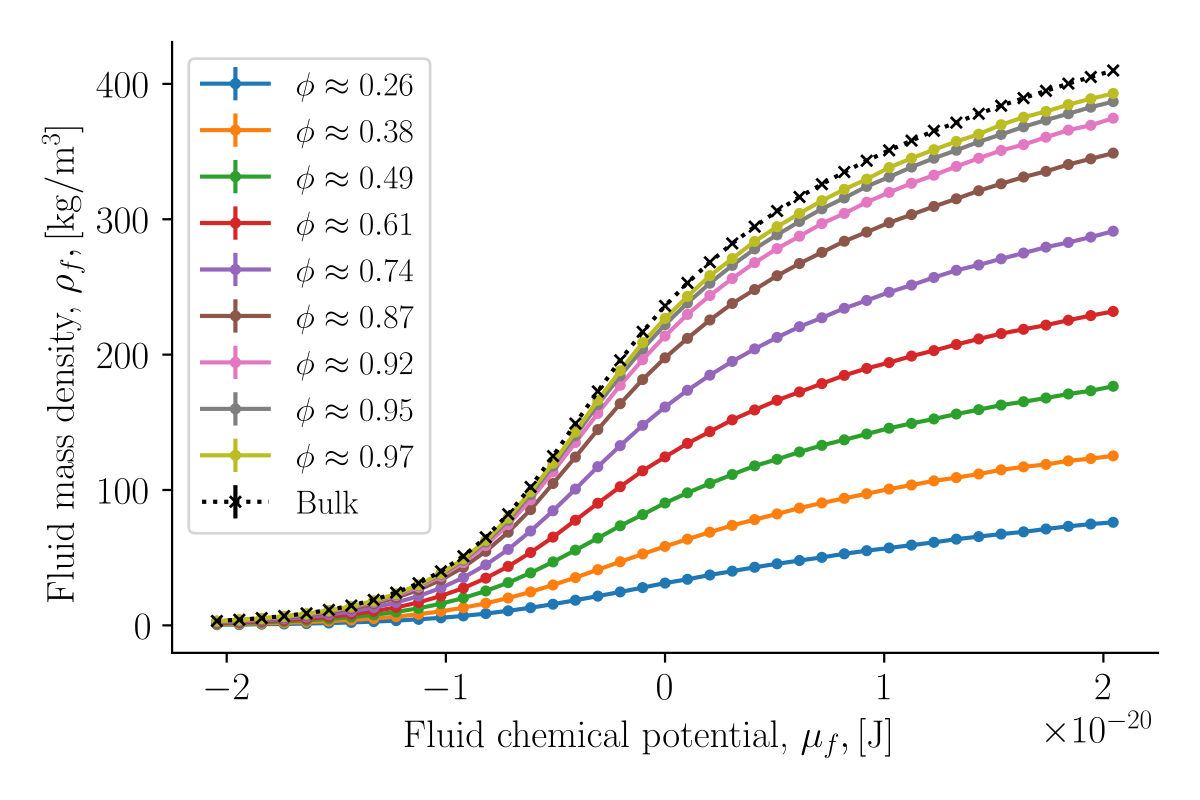}
    \caption{Fluid mass density as a function of fluid chemical potential for varying porosities. Bulk fluid values are indicated by black crosses at the top.}
    \label{fig:fluid_density}
\end{figure}

The fluid mass density in the REV is shown as a function of fluid chemical potential in Fig. \ref{fig:fluid_density}. The chemical potential was set by the environment. The fluid mass density is shown for a varying porosity. We see that the fluid mass density in the REV increases monotonically with the chemical potential, approaching the value of the adjacent bulk phase  (shown by black crosses in the figure). This approach is as expected.

The fluid mass density of the REV differs from the mass density outside the porous medium. But, the fluid mass density of the REV, corrected by the porosity will also differ from the outside value (not shown in the figure). This indicates that thermodynamic properties change upon fluid confinement. In the low porosity case, the effect is large. For a fluid chemical potential near \SI{0}{\joule}, the density $\rho_f$ changes by a factor of 50, as the porosity changes from densest packing of spheres to bulk fluid. Every single curve deviates from the bulk vapor mass density when the fluid chemical potential is around $ \SI{-e20}{\joule}$, where the adsorption of fluid into the pores starts to seriously increase. The numbers refer to the chosen model; methane. 

\begin{figure}
    \centering
    \includegraphics[width=0.8\linewidth]{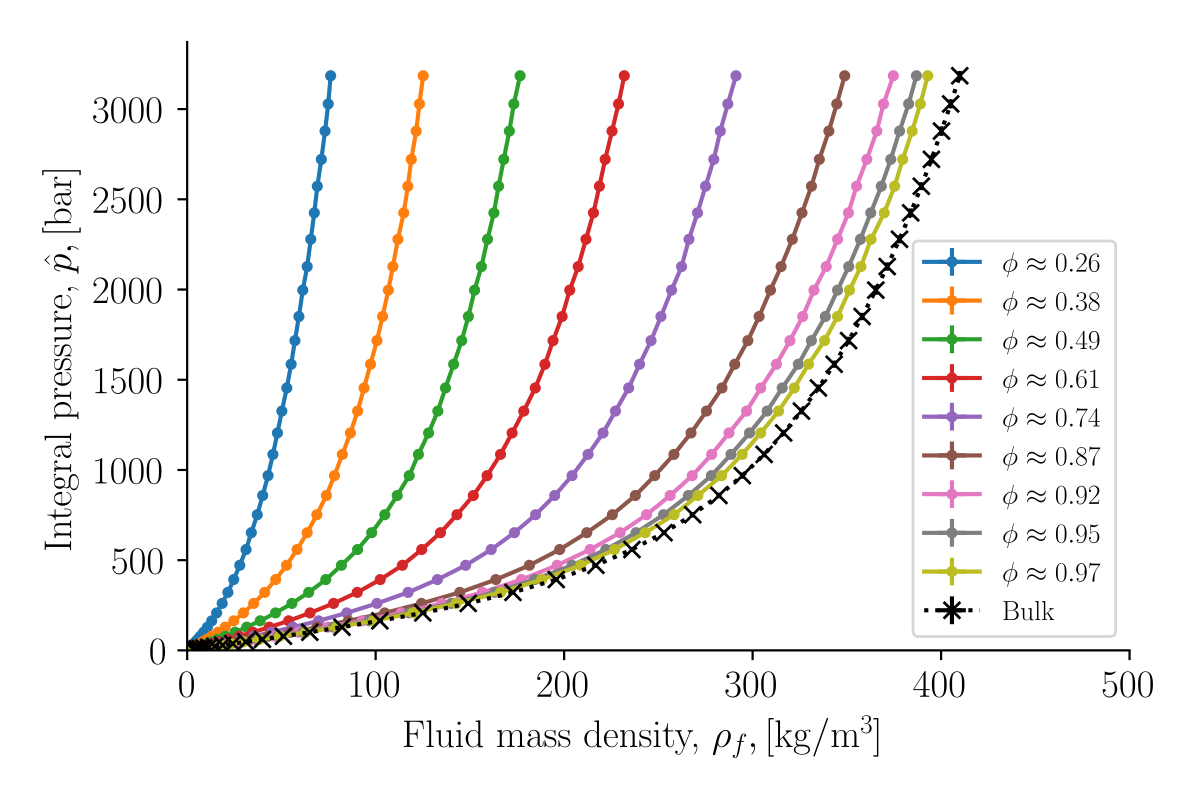}
    \caption{Integral pressure as a function of fluid mass density for varying porosities, or equation of state for the porous medium. Bulk fluid values are shown by black crosses at the bottom.}
    \label{fig:integral_pressure_fluid_density}
\end{figure}

The dependence of the integral pressure on the fluid mass density for the various porosities is presented in Fig. \ref{fig:integral_pressure_fluid_density}. The rise in the integral pressure from a few bars to $\SI{3184(27)}{\bar}$ is shown for all porosities. The slope of the curves increases with decreasing porosity. The bulk isotherm is again indicated by black crosses and gives the lower limit of the variation. Again, the curves approach the bulk value as the porosity increases, as expected. The curves apply for a temperature $T=2.0\epsilon/k_\text{B}=\SI{296.2}{\kelvin}$. 

The mean free path of the bulk vapor densities can be estimated with the kinetic theory of gases with the equation $\ell=(\sqrt{2}\pi\sigma^2n_b)^{-1}$, where $n_b$ is the number density of fluid particles in the bulk phase. For the lowest bulk mass density, $\rho_b=\SI{3.3(2)}{\kilo\gram\per\meter}$ the mean free path is estimated to be $\ell=\SI{12.7(9)}{\nano\meter}$. The octahedral voids in the fcc lattice can accommodate spheres with diameter \SI{1.6}{\nano\meter} to \SI{11.4}{\nano\meter} for the varying porosities, in which case the system is in the Knudsen flow regime. For a bulk mass density $\rho_b=\SI{24.2(6)}{\kilo\gram\per\meter}$, the mean free path is estimated to be $\ell=\SI{1.71(5)}{\nano\meter}$. The system will be in the Knudsen flow regime for the lowest densities.

The curves obtained for the equilibrium condition Eq. \ref{eq:equi}, can be regarded as an equation of state relating the temperature, porosity, and fluid mass density of the REV to the integral pressure, see equation Eq. \ref{eq:EoS}. we shall use the sets of relations in the same manner as an equation of state. Once we know the fluid mass density of the REV, we know also the chemical potential that controls it from Fig. \ref{fig:fluid_density}, and therefore also the corresponding integral pressure from Fig. \ref{fig:integral_pressure_fluid_density}.  
 
\subsection{Non-equilibrium conditions}

By applying the reflective boundary method we generated a mass flux, a fluid mass density difference, and a difference in integral pressure difference across the porous medium. By increasing the reflecting probability $q$ we increased the gradients. In this manner, we varied the integral pressure gradient between approximately $\Delta \hat{p}/L = \SI{-7}{\bar\per\micro\meter}$ and \SI{20000}{\bar\per\micro\meter}. The corresponding Reynolds numbers varied from approximately zero up to $\text{Re} = 3.54\pm0.08$. Gradients that are generated in non-equilibrium molecular dynamics can be very large because the length of the simulation box is relatively short and the method can apply strong forces.

\begin{figure}
    \centering
    \includegraphics[width=0.8\linewidth]{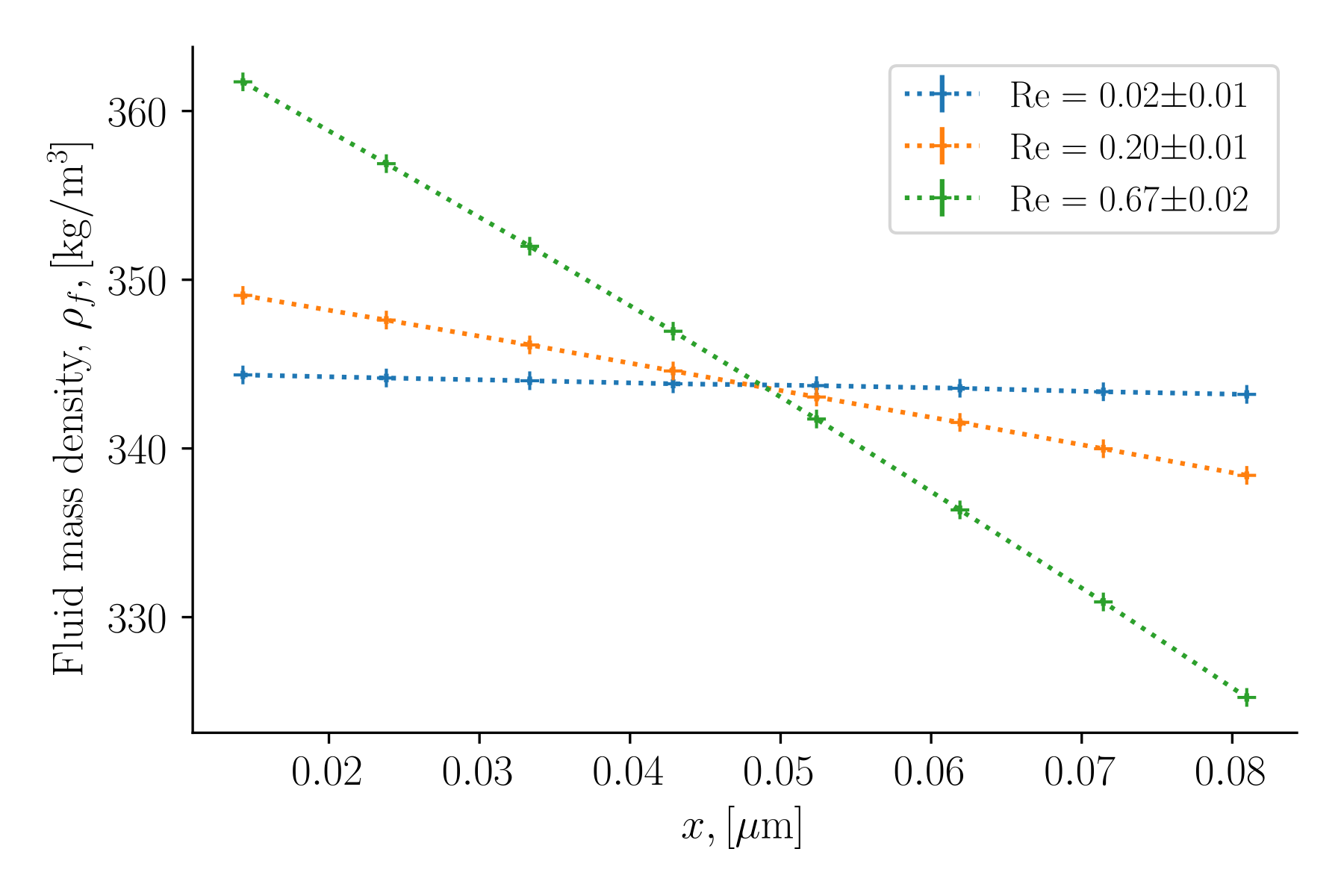}
    \caption{Fluid mass density as a function of the $x$-coordinate for porosity $\phi \approx 0.87$, average fluid mass density $\rho_b=\SI{351.3(7)}{\kilo\gram\per\meter^3}$, and varying Reynolds numbers.}
    \label{fig:fluid_mass_density_profile}
\end{figure}

\begin{figure}
    \centering
    \includegraphics[width=0.8\linewidth]{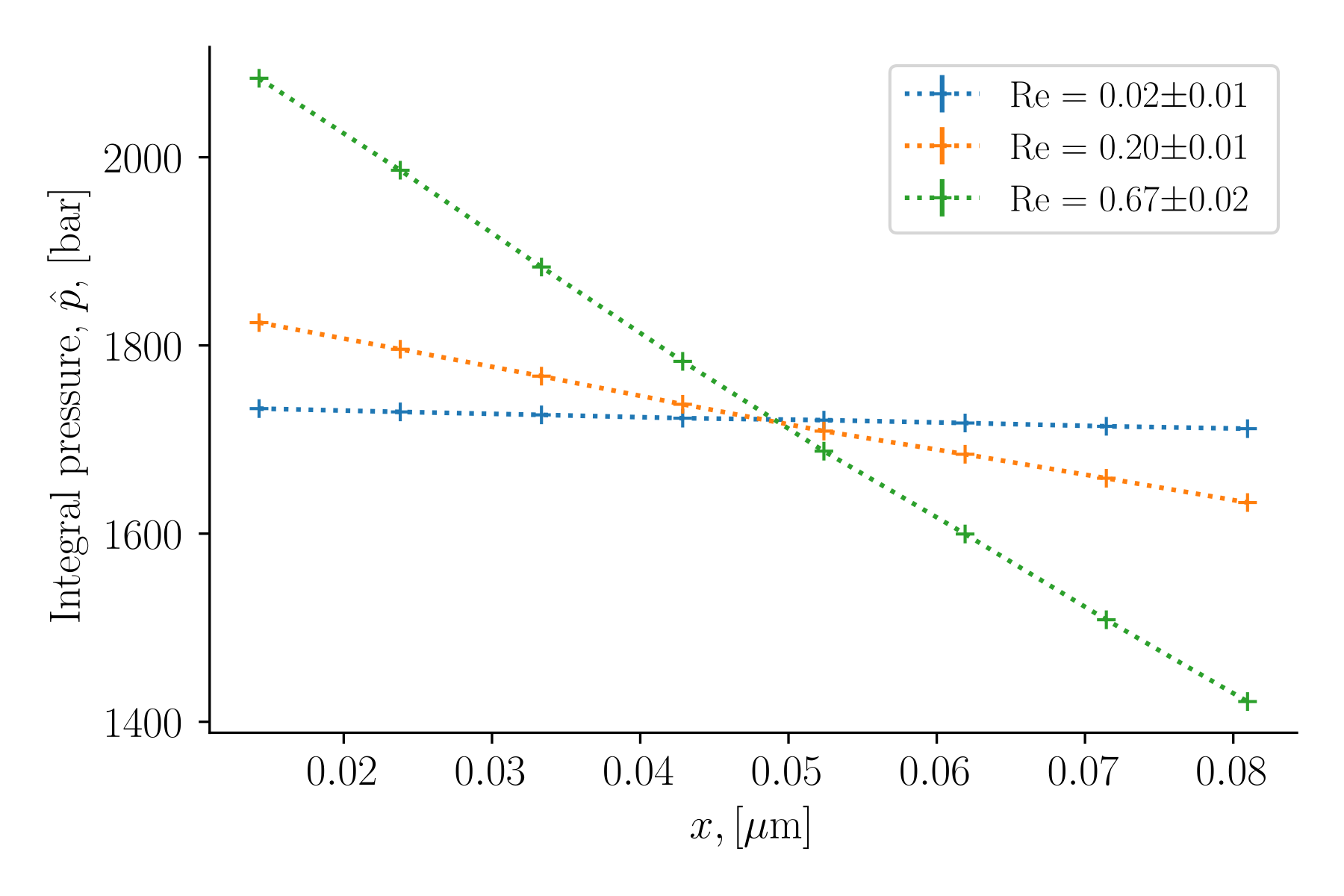}
    \caption{Integral pressure as a function of the $x$-coordinate for porosity $\phi \approx 0.87$, average fluid mass density $\rho_b=\SI{351.3(7)}{\kilo\gram\per\meter^3}$, and varying Reynolds numbers.}
    \label{fig:pressure_profile}
\end{figure}

The fluid mass density profiles across the porous medium are shown in Fig. \ref{fig:fluid_mass_density_profile} for a porosity $\phi \approx 0.87$, and an average fluid mass density of the REV $\rho_f = \SI{351.3( 7)}{\kilo\gram\per\meter^3}$. The Reynolds numbers varied from $\text{Re}=0.02\pm0.01$ to $\text{Re}=0.67\pm0.02$. From the fluid mass density profiles, we obtained the integral pressure profiles. Examples are shown in Fig. \ref{fig:pressure_profile}. The integral pressure gradients were calculated from the integral pressure profiles. The gradient in integral pressure was in good approximation constant. This was not expected and is also not needed in the data reduction procedures. Because the mass flux is constant at a steady state (mass conservation), and the driving force is approximately constant, it follows that $\rho_f l$ and $\rho_f k/\eta $ were constant across the porous medium. The conductivity and the permeability are not necessarily constant in a porous medium of varying porosity. 

\begin{figure}
    \centering
    \includegraphics[width=0.8\linewidth]{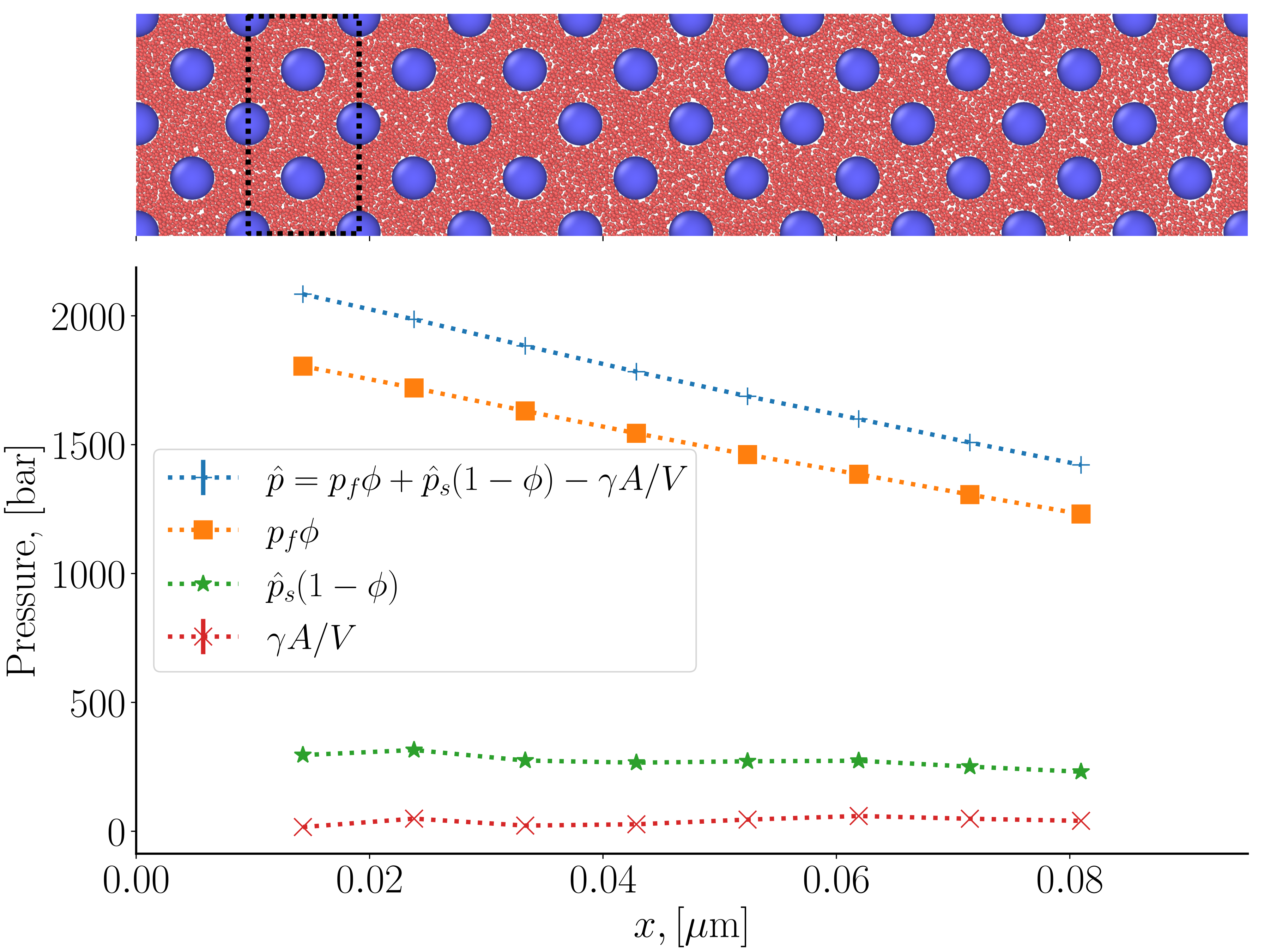}
    \caption{Profile of porosity $\phi \approx 0.87$, average fluid mass density $\rho_b=\SI{351.3(7)}{\kilo\gram\per\meter^3}$ and Reynolds number $\text{Re}=0.67\pm0.02$. \textbf{Top:} Visualization of the simulation box. The border of a layer of volume $V_l$ is marked with a dashed line. \textbf{Bottom:} Contributions from the bulk phases and the surface to the integral pressure as function of the $x$-coordinate. The $x$-axis of the visualization correspond to the graph. The points in the graph gives the $x$-position of the center of the REV (the volume $V_l$).}
    \label{fig:pressure_contributions_profile}
\end{figure}

There is not yet much experience with the integral pressure of porous media reported in the literature. It is therefore appealing to examine the contributions to the integral pressure from the bulk phases and surface, and test our way to compute $\hat{p}$ from the equation of state, Eq. \ref{eq:EoS}. By assuming $\hat{p}_f = p_f$, $\hat{p}=p_f$ and $\hat{\gamma}=\gamma$, the individual contributions can be calculated. The outcome is illustrated in Fig. \ref{fig:pressure_contributions_profile}, again for a porosity $\phi\approx0.87$, average fluid mass density $\rho_b=\SI{351.3(7)}{\kilo\gram\per\meter^3}$. The Reynolds number was now $\text{Re}=0.67\pm0.02$. The assumptions hold for large porosities when the disjoining pressure is zero, and when the surface tension is independent of the fluid-solid surface curvature. These assumptions are not necessary for the continued analysis of this work, they are just chosen to illustrate a case with individual contributions. The surface tension $\gamma$ was calculated from the spherical mechanical pressure tensor. The value increases monotonically with increasing density between \SI{0.079(9)}{\milli\newton\per\meter} and \SI{34(1)}{\milli\newton\per\meter}. See the supplementary information for more details on the surface tension.

The integral solid pressure was calculated from the integral pressure and surface tension
\begin{equation}
    \hat{p}_s = p_f + \gamma A/V_s
\end{equation}
where $A$ is the fluid-solid surface area and $V_s$ is the volume of the solid phase, see Eq. \ref{eq:grand}. In Fig. \ref{fig:pressure_contributions_profile}, we see the local contributions from the gradient in fluid pressure $\phi \nabla p_f = \SI{-8629(106)}{\bar\per\micro\meter}$, in integral solid pressure $(1-\phi) \nabla\hat{p}_s = \SI{-9965(215)}{\bar\per\micro\meter}$, and in surface tension $(A/V)\Delta\gamma = \SI{370(214)}{\bar\per\micro\meter}$. The single parts sum to the total value shown in the figure. This value of $\hat{p}$ was also determined from the equation of state, Eq. \ref{eq:EoS}. Within the accuracy of the calculation, we confirmed our hypothesis that the two routes give the same result. 

A control of isothermal conditions was carried out. We know that a pressure difference may generate a temperature difference, or vice versa \cite{rauter2021thermo}. Constant temperature is therefore a condition for the single flux-force product of the entropy production, and consequently for Darcy's law's applicability. We, therefore, confirmed that the temperature was constant across the system.

\begin{figure}
    \centering
    \includegraphics[width=0.8\linewidth]{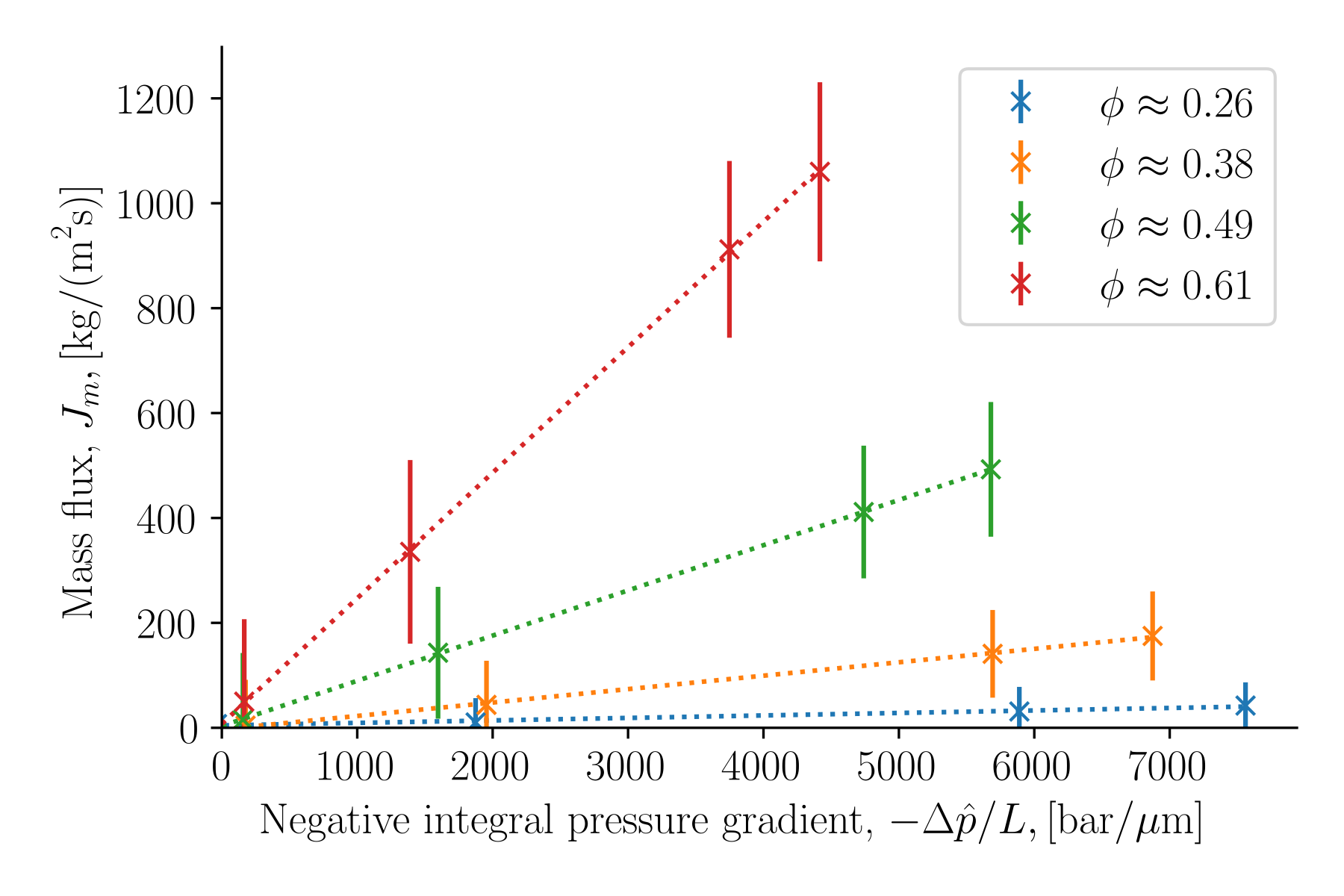}
    \caption{The mass flux as a function of the negative integral pressure gradient for porosities $\phi\approx[0.26,0.61]$ and average bulk fluid mass density $\rho_b = \SI{108(1)}{\kilo\gram\per\meter^3}$. The dashed lines are linear fits to the points.}
    \label{fig:flux_force}
\end{figure}

Figure \ref{fig:flux_force} shows plots that are used to determine the conductivity and permeability in Darcy's law. Rather than using the volume flux, we have used the mass flux on the ordinate axis, as this flux, but not the volume flux is constant across the porous media in a steady state. The figure illustrates varying porosities $\phi\approx [0.26, 0.61]$ and the average bulk fluid mass density was $\rho_b =\SI{ 108(1)}{\kilo\gram\per\meter^3}$ in these plots.

We see that the mass fluxes in all cases can be regarded as linear functions of the negative integral pressure gradient; the dashed lines are linear fits to the calculated mass fluxes. Also observed is that the fluxes intersect the origin of the axes. This behavior is compatible with a linear theory like NET. There is no threshold for transport at low-pressure gradients. 

\subsection{Conductivity and permeability}

\begin{figure}
    \centering
    \includegraphics[width=0.8\linewidth]{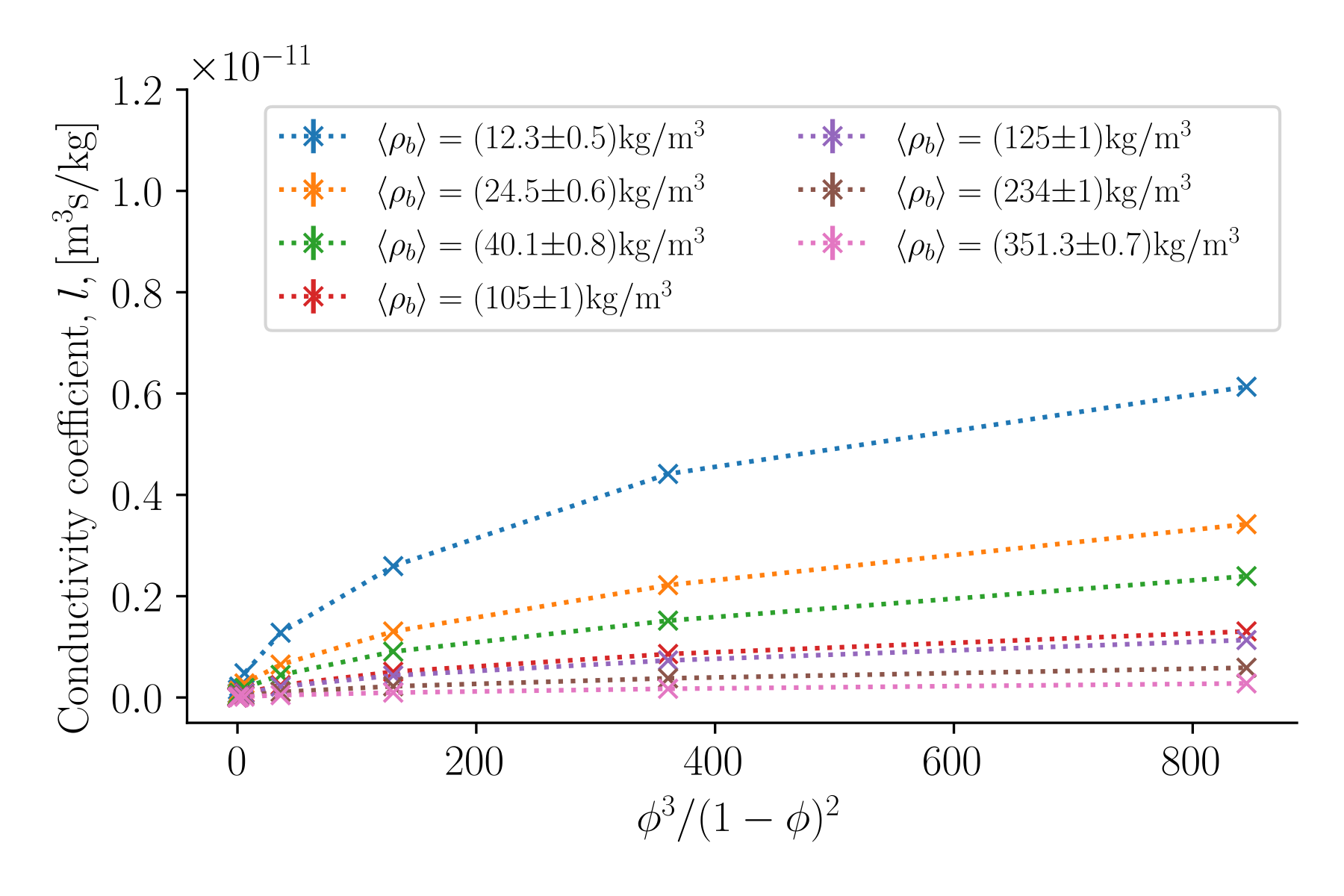}
    \caption{Conductivity coefficient as a function of the porosity factor $\phi^3/(1-\phi)^3$ for various average fluid mass densities.}
    \label{fig:conductivity_porosity}
\end{figure}

Fig. \ref{fig:pressure_profile} gave the integral pressure gradient inside the porous medium. The integral pressure gradient from this figure together with the constant mass flux was used to determine the conductivity of the porous medium. The conductivity was plotted in Fig. \ref{fig:conductivity_porosity} as a function of the factor $\phi^3/(1-\phi)^2$ in the Kozeny-Carman equation. There was no convincing relationship between the variables.

\begin{figure}
    \centering
    \includegraphics[width=0.8\linewidth]{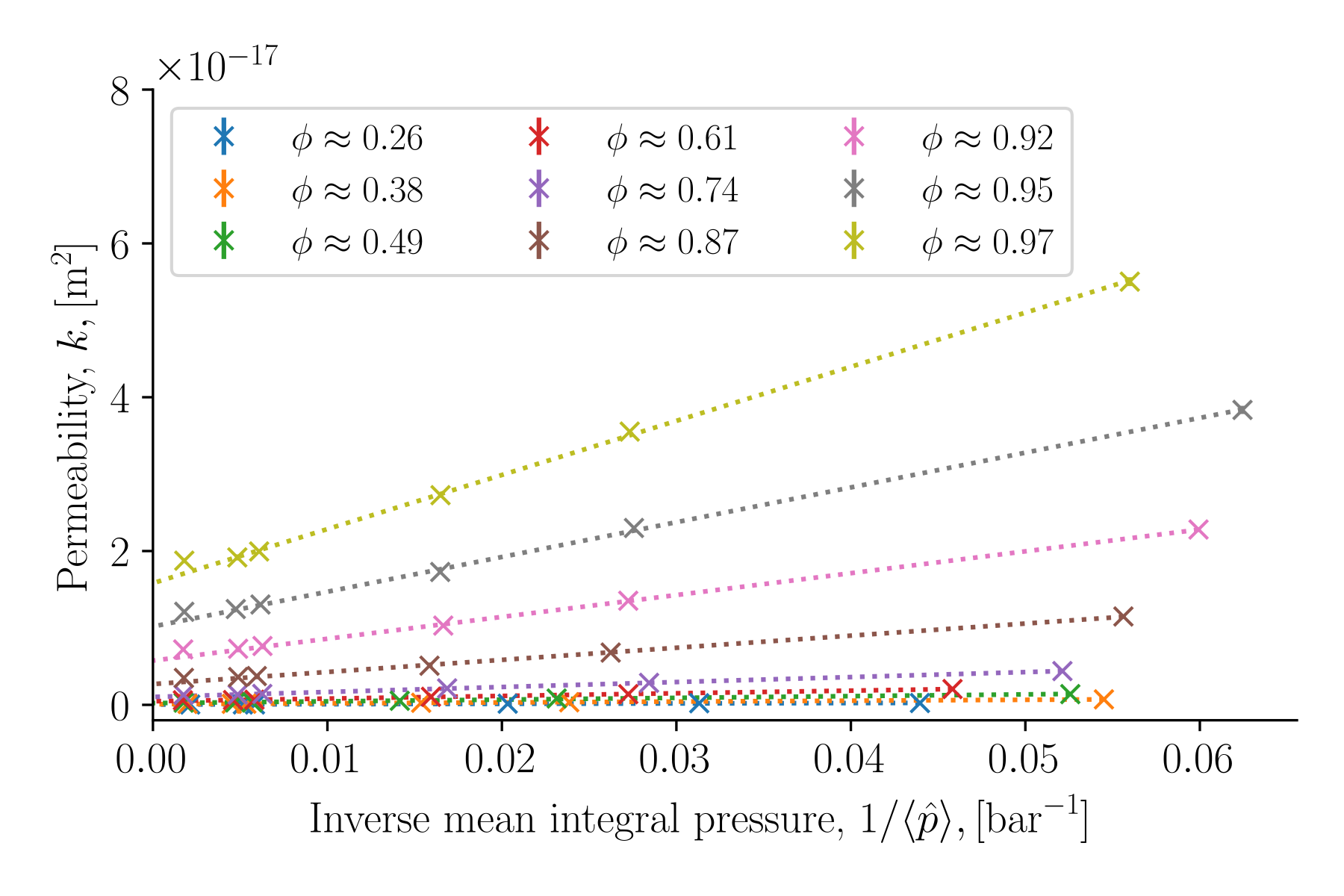}
    \caption{Permeability as a function of inverse average pressure pressure.}
    \label{fig:permeability_pressure}
\end{figure}

It is more interesting to consider the permeability $k$. Its value was computed from the conductivity $l$ and the shear viscosity $\eta_b$, and the results are shown as a function of the inverse average integral pressure (which here happens to be the same as the inverse differential pressure) in Fig. \ref{fig:permeability_pressure}. We see that the permeability increases with a constant slope for large values of the inverse average integral pressure (low densities), where the fluid is highly compressible. For small values (high densities), the slope decreases as the fluid become less compressible. This behavior is expected from the Klinkenberg correction \cite{Klinkenberg1941,Baehr1991}. The results at the low pressure were therefore fitted to a straight line and extrapolated to the high-pressure end to obtain the absolute permeability, $k_0$.

\begin{figure}
    \centering
    \includegraphics[width=0.8\linewidth]{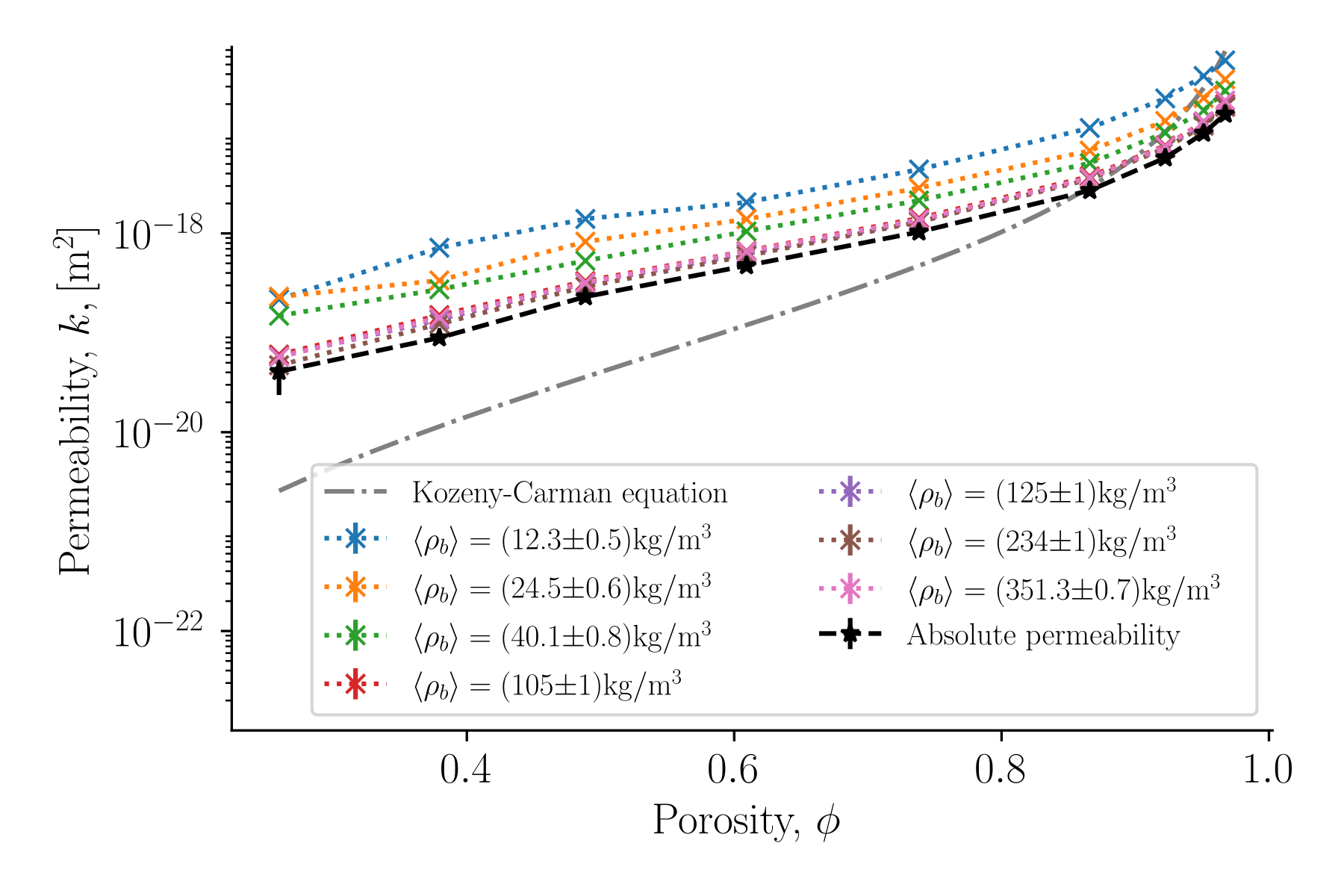}
    \caption{Permeability in logarithmic scale as a function of the porosity $\phi$ for varying average bulk fluid mass density. The Kozeny-Carman equation is shown as a black dashed curve for a tortuosity $\tau=1$.}
    \label{fig:log_permeability}
\end{figure}

The dependence of the permeability on other lattice parameters is shown in the log-plot of the permeability as a function of the porosity in Fig. \ref{fig:log_permeability}. A systematic variation is demonstrated over a change in four orders of magnitude of $k$. The absolute permeability was calculated using the Klinkenberg correction formula (in the range where this correction applies), by plotting the permeability vs $1/p$ and extrapolating to very large pressures. The absolute permeability is shown as a black dashed line. We see that the absolute permeabilities form a lower limit for the family of calculated permeabilities. On the other hand, the Kozeny-Carman equation with tortuosity $\tau=1$, shown as a grey dash-dotted line, does not bring out any new physical insight.

\begin{figure}
    \centering
    \includegraphics[width=0.8\linewidth]{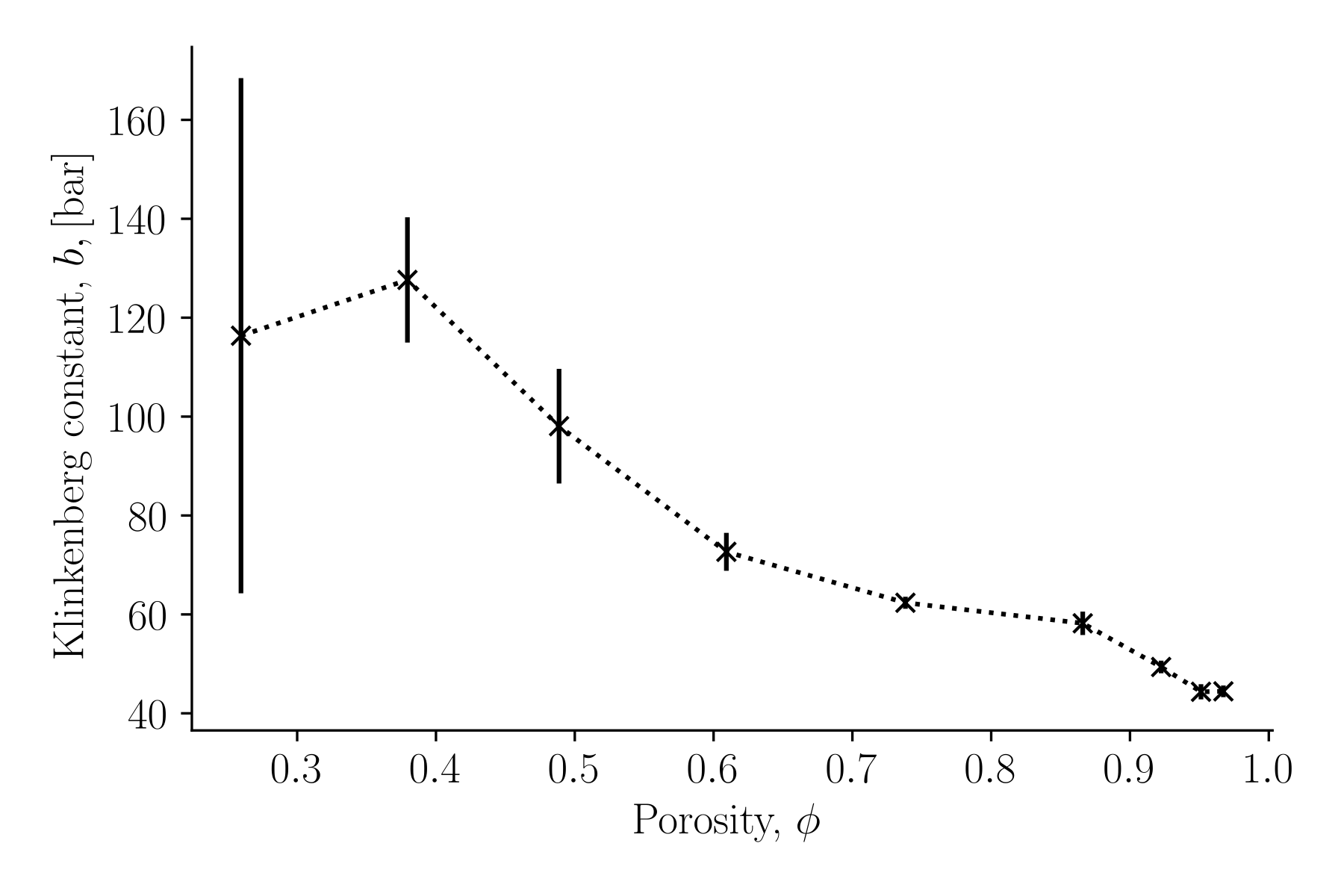}
    \caption{Klinkenberg constant $b$ as a function of porosity $\phi$.}
    \label{fig:klinkenberg}
\end{figure}

Fig. \ref{fig:klinkenberg} shows the Klinkenberg coefficient $b$ as a function of porosity $\phi$. The coefficient is porosity dependent and significant for the whole range of porosities. To a good approximation, the dependence is linear for porosities below 0.6. The coefficient can be understood from the lack of interaction between solid spheres and fluid particles on the particle level, which here is consistent with slippage.

\section{Conclusion and Perspectives}

We have demonstrated a new thermodynamic procedure to find the local effective pressure gradient that drives the mass flow through a porous medium. The REV was constructed from additive thermodynamic variables, such that the Gibbs equation and in turn the entropy production and flux-force equations could be derived. The method was applied to single-phase fluid flow in an isothermal medium of varying porosity and fluid mass density under laminar flow conditions. The hydraulic conductivity as well as the permeability were shown to vary with porosity and fluid mass density, and give values that are typical in the literature. The systems studied in this work were in steady-state. This does not pose any limitation on the method. The procedure can be used to describe transients. It is interesting that the system supported the behavior behind the Klinkenberg effect, and that we obtained variables of a size compatible with this effect, \textit{e.g.} $k < 10^{-13}$m$^2$. 

The procedure has its thermodynamic basis in standard non-equilibrium thermodynamics as combined with Hill's method of nanothermodynamics to describe the confined fluid. This method should now be tested with a two-phase flow, to help solve problems stated in the literature on the upscaling problem; i.e. on how we can properly describe the porous medium microstates, that is the origin of Darcy scale behavior. For instance, we can define the effective pressure without consideration of the capillary pressure. For two-phase flow, the REV will be larger than a unit cell to include the statistical variation of the two-phases.

A particular symmetric lattice was chosen to illustrate the derivations. This should also not be regarded as a limitation. The method could be accommodated to deal with pore distributions. It can be applied to highly confined systems, where the disjoining pressure is significant. The central point is the proper construction of the grand potential.

\vspace{0.25cm}
\noindent\textbf{Competing Interests}
The authors have no relevant financial or non-financial interests to disclose.

\vspace{0.25cm}
\noindent\textbf{Author Contributions}
O.G. contributed to formal analysis, investigation, methodology, software, visualization, conceptualization, writing original drafts, reviewing, and editing. S.K. and D.B. contributed to formal analysis, investigation, methodology, conceptualization, supervision, writing original drafts, reviewing, and editing. T.T.T. computed viscosities and contributed to reviewing and editing. M.S.B. and M.T.R. contributed to early versions of software, visualization, methodology, conceptualization, reviewing, and editing. All authors have read and agreed to the published version of the manuscript.

\vspace{0.25cm}
\noindent\textbf{Funding}
This work was funded by the Research Council of Norway through its Centres of Excellence funding scheme, project number 262644, PoreLab.

\vspace{0.25cm}
\noindent\textbf{Acknowledgments}
The authors acknowledge valuable comments to the last version of the manuscript from Steffen Berg and Hamidreza Erfani Gahrooei. The simulations were performed on resources provided by UNINETT Sigma2 - the National Infrastructure for High-Performance Computing and Data Storage in Norway with project numbers nn9229k and nn8022k. We thank the Research Council of Norway for its Centres of Excellence funding scheme, project number 262644, PoreLab.

\pagebreak
\vspace{2cm}
\begin{center}
 {{\LARGE \center Supplementary information to "Local thermodynamic description of isothermal single-phase flow in porous media"}}\\
{{\large\center Olav Galteland, Michael T. Rauter, Mina S. Bratvold, Thuat T. Trinh, Dick Bedeaux and Signe Kjelstrup}}\\
{{\large\center PoreLab, Department of Chemistry, Norwegian University of Science and Technology}}\\
{{\large\center March 4 2022}}
\end{center}
\vspace{1cm}

\setcounter{section}{0}
\section{Surface tension}
\begin{figure}
    \centering
    \includegraphics[width=0.8\linewidth]{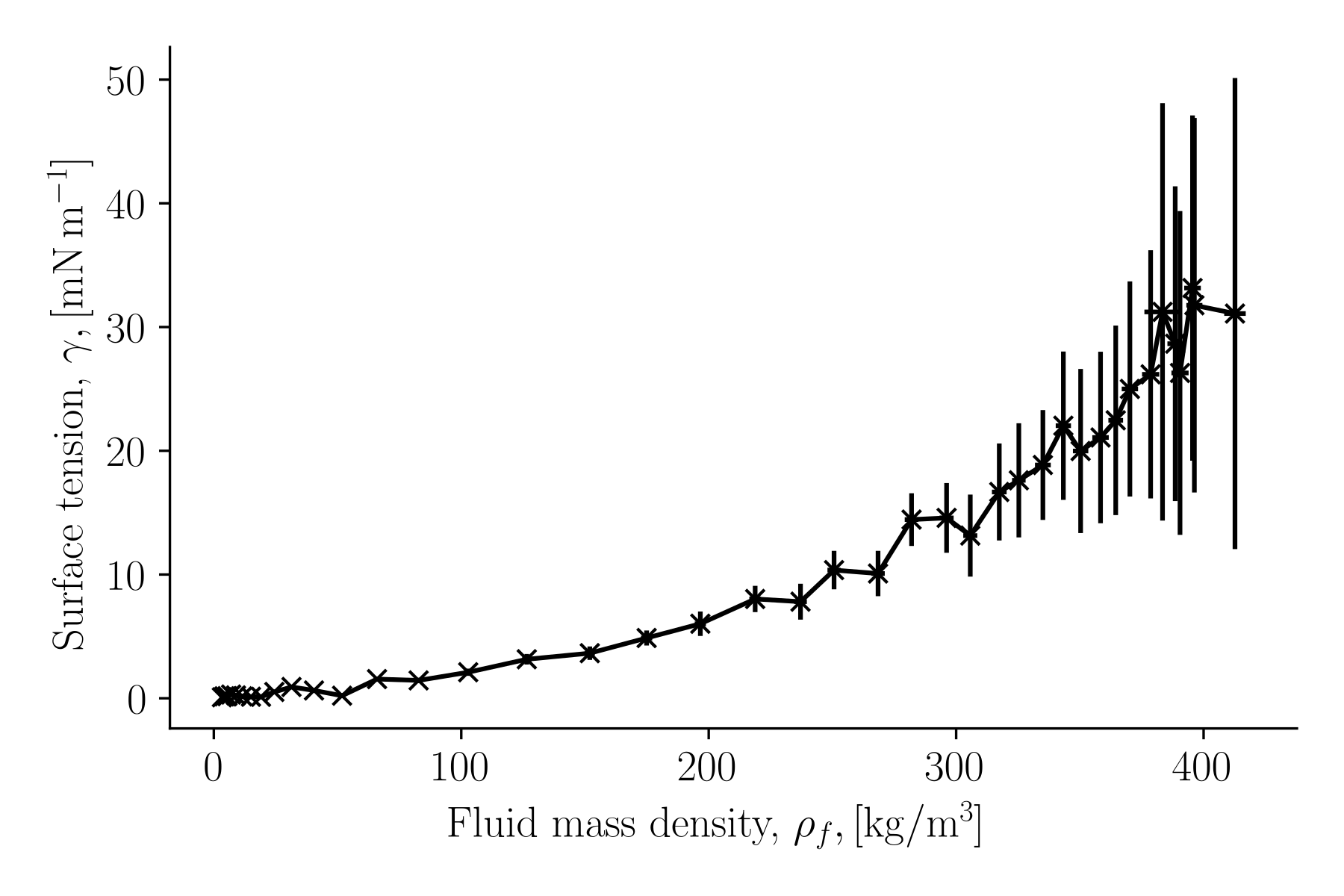}
    \caption{Solid-fluid surface tension as a function of fluid mass density for a single solid sphere surrounded by a fluid.}
    \label{fig:surften}
\end{figure}

The fluid-solid surface tension was calculated from the spherical mechanical pressure tensor of a single solid sphere surrounded by a fluid of varying mass density. This was considered as the surface tension of a lattice with a large lattice constant, such that the solid surfaces were far apart. The mechanical pressure tensor can be written as a sum of an ideal gas contribution and a virial contribution. The diagonal components of the mechanical pressure tensor were calculated in spherical shells with origin at the center of the solid sphere
\begin{equation}
    P_{\alpha\beta}(r) = \rho(r) k_\text{B}T\delta_{\alpha\beta}+P_{\alpha\beta}^v(r),
\end{equation}
where $r$ is the distance to the origin, $\rho(r)$ is the fluid mass density of the spherical shell, $\delta_{\alpha\beta}$ is the Kroenecker delta and $P^{s,v}_{\alpha\beta}$ is the virial contribution. The subscripts $\alpha$ and $\beta$ refer to the spherical coordinates, $(r, \theta, \phi)$. The virial contribution is
\begin{equation}
    P_{\alpha\beta}^v(r) = - \frac{1}{V(r)}\sum_{i=1}^N\sum_{j>i}^Nf_{ij,\alpha}\int_{C_{ij}\in V(r)}\text{d} l_\beta.
\end{equation}
where $V(r)$ is the volume of the spherical shell, $f_{ij,\alpha}$ is the $\alpha$-component of the force acting on particle $i$ due to particle $j$. The line integral is along the Irving-Kirkwood contour $C_{ij}$ \cite{Irving1950, Harasima1958, Schofield1982}, which is the straight line between particle $i$ and $j$. The line integral gives the length of the $\beta$-component of the contour $C_{ij}$ that is contained in the spherical shell $V(r)$. The surface tension is calculated from the pressure tensor
\begin{equation}
    \gamma = \frac{1}{R^2}\int_R^{r_0}(P_N-P_T) r^2\text{d} r,
    \label{eq:surface_tension}
\end{equation}
where $P_N = P_{rr}$ is the normal to the fluid-solid surface and $P_T=(P_{\theta\theta}+P_{\phi\phi})/2$ is tangential to it. The integral limit $r_0$ is a position in the fluid far away from the fluid-solid surface such that $P_N=P_T$. The surface tension is shown as a function of fluid mass density in Fig. \ref{fig:surften}.

\bibliographystyle{ieeetr}
\bibliography{library}
\end{document}